\documentstyle[prl,aps,preprint,tighten]{revtex}
\begin{document}
\draft
\title{Intensity distribution of waves transmitted through a multiple
scattering medium}
\author{Th.~M.~Nieuwenhuizen and M.~C.~W. van Rossum}
\address{Van der Waals-Zeeman Laboratorium, Universiteit van Amsterdam
\\ Valckenierstraat 65, 1018 XE Amsterdam, The Netherlands}
\date{\today}\maketitle
\begin{abstract}
The distributions of the angular transmission coefficient and of the total
transmission are calculated for multiple scattered waves. The calculation is
based on a mapping to the distribution of eigenvalues of the transmission
matrix. The distributions depend on the profile of the incoming beam. The
distribution function of the angular transmission has a stretched exponential
decay. The total-transmission distribution grows log-normally whereas it
decays exponentially. \end{abstract}

\preprint
\pacs{42.25-p,78.20.Dj,72.10.-d,05.40+j}
\date{today}

Transport through mesoscopic systems is a wide field of interest and is
studied with a variety of waves: sound, microwaves, electrons, and light. The
wave character leads to interference between the transmission channels. This
causes large fluctuations. Therefore the observables are not always
characterized by mean values, but their entire distribution functions are of
interest. This is particularly prominent in the distribution of eigenvalues
of the transmission matrix. The eigenvalues have a bimodal distribution
peaked around 0 and 1 with a mean value $\ell/ L \ll 1$, where $\ell$ denotes
the mean free path and $L$ the thickness of the
sample\cite{dorokhov,pendry,stoneboek}.

Three distinct transmission quantities can be measured. If a laser
illuminates a diffuse medium the transmitted signal consists of speckles.
These speckles are the first quantity of interest. The intensities of the
speckles obey in first approximation the Rayleigh law, which is a negative
exponential distribution. In the mesoscopic regime interference modifies this
distribution. The leading correction was derived by Shnerb and Kaveh
\cite{shnerb}. Genack and Garcia observed a deviation from the Rayleigh
law at large intensities\cite{genack2}. A crossover to stretched exponential
behavior was derived by Kogan {\it et al.} \cite{kogan}.

The second quantity is the total transmission. It is obtained by integrating
over the outgoing surface, thus collecting all speckles. When adding the
speckles of the $N$ independent channels, the law of large numbers predicts a
Gaussian distribution with variance of the order $1/N$. However, interference
broadens the variance to $L/(\ell N)$ \cite{stephen1}. This was confirmed in
a recent experiment of de Boer {\it et al.}\cite{t3prl}. Also in computer
simulations by Edrei {\it et al.} the Gaussian behavior was observed, while a
cross over to a log-normal distribution was seen for large
disorder\cite{edrei}. De Boer {\it et al.} also measured a small but clear
deviation from a Gaussian, due interference of three transmission channels.

The third quantity which can be extracted is the conductance, which is
obtained by summing also over all incoming directions.
An extensive discussion of the distribution function of the
conductance including a review was given by Altshuler {\it et
al.}\cite{altshulerboek}.

These three quantities are dominated by different interference processes.
This can already be seen by considering their autocorrelation functions
\cite{feng,mello}. Let $T_{ab}$ be the angular transmission coefficient for a
speckle spot $b$ arising from incoming wave $a$. Its correlator is called the
$C_1$ correlation function. It is dominated by a {\em disconnected} diagram.
The correlator of the total transmission $T_a=\sum_bT_{ab}$ is called the
$C_2$ correlation, and dominated by the interference of two diffusons. In a
diagrammatic approach it is a {\em connected, loopless} diagram. The
interference process mixes two incoming into two outgoing diffusons.
Finally, the correlator of the conductivity $T=\sum_{ab}T_{ab}$ is called
$C_3$ correlation or Universal Conductance Fluctuation. It is dominated by a
{\em connected loop} diagram.

We shall first calculate the distribution function of the total
transmission.
Subsequently we apply these ideas to the statistics of the angular
transmission coefficient.

Consider a three dimensional slab with dimensions $W\times W\times L$, ($W\gg
L$) with elastic scatterers at quenched random positions. We consider the
`metallic' regime, where the dimensionless conductance $g$ is large. Let us
look at the transmission problem for a
scalar plane wave with unit incoming flux in channel $a$,
\begin{equation}
\psi_a^{\rm in}({\bf
r})=\frac{1}{\sqrt{Ak \mu_a}}\exp(i{\bf q}_a\rho+ik \mu_a z) \qquad z<0,
\label{planepsi}\end{equation}
where $A=W \times W $ denotes the area of the slab, $k$ is the wavenumber,
$\rho=(x,y)$ is the transversal coordinate and $\mu_a=\sqrt{1-{\bf
q}_a^2/k^2}=\cos\theta_a$, where $\theta_a$ is the angle with respect to the
$z-$axis. The precise form of the propagators was derived in Ref.\
\cite{thmn5}. An incoming diffuson has the form
\begin{equation} {\cal L}_a^{\rm in}({\bf r})=\frac{4\pi \tau_1(\mu_a)}
{k\ell A\mu_a}\,\frac{L-z}{L}, \end{equation} where $\tau_1$
describes the limit intensity of a semi-infinite system. We shall term the
propagator summed over all incoming channels the {\em `total-flux diffuson'}:
\begin{equation}\label{fluxladder}
{\cal L}^{\rm in}({\bf r})\equiv \sum_a {\cal L}^{\rm in}_a({\bf r}) =
\frac{4k}{\ell} \frac{L-z}{L}.\end{equation} In the bulk the difference
between a total-flux diffuson and a diffuson arising from plane wave
incidence is just described by an overall factor $\epsilon_a$,
\begin{equation} {\cal L}_a^{\rm in}({\bf r})=\epsilon_a{\cal L}^{\rm
in}({\bf r}),  \label{fluxladder2}\end{equation} with $ \epsilon_a \equiv
\tau_1(\mu_a)\pi /(k^2A\mu_a) $ satisfying the sum rule
$\sum_a\epsilon_a=1$. Similarly, the outgoing diffusons are\begin{equation}
{\cal L}_b^{\rm out}({\bf r})=\epsilon_b{\cal L}^{\rm out}({\bf r}),\qquad
{\cal L}^{\rm out} ({\bf r})=\frac{k}{\ell} \frac{z}{L}.\end{equation}

The wave is transmitted into outgoing channel $b$ with transmission amplitude
$t_{ab}=2k\sqrt{\mu_a \mu_b} G_{ab}$ and transmission probability $T_{ab}
\equiv |t_{ab}|^2$. The average total transmission is obtained
by summing all outgoing channels \begin{equation} \langle T_a\rangle=\langle
\sum_{b}T_{ab}\rangle = \frac{\tau_1(\mu_a)\ell}{3L\mu_a}.\end{equation} The
average conductance is given by  \begin{equation} g\equiv \sum_a \langle
T_a\rangle=\frac{k^2 A\ell}{3\pi L},\end{equation} thus $\langle T_a\rangle
=\epsilon_a g$, while one also has $\langle T_{ab}\rangle= \epsilon_a
\epsilon_b g$ \cite{thmn5}.

We consider the $j^{\rm th}$ cumulant of $T_a$. In a diagrammatic
approach this object has $j$ transmission amplitudes $t_{ab}$ and an equal
number of Hermitean conjugates $t_{ba}^\dagger=t_{ab}^*$. Explicit
calculation for
the second and third cumulant showed that the leading diagrams are
connected, but have no loops\cite{t3prl}. We assume that this holds to
any order. Let us fix the external diffusons in the term
$t_{a b_1}t^\dagger_{b_1a}t_{ab_2}\cdots t_{ab_j}t^\dagger_{b_ja}$.
Contributions to the sum over $b_i$ only come from diagrams with
outgoing diffusons that have no transversal momentum.
These are the diagrams where the lines with equal $b_i$ are paired
into diffusons. The outgoing diffusons are fixed now. For connected
loopless diagrams
there are $(j-1)!$ distinct choices for pairing the incoming
amplitudes.
Next we factor out the incoming and outgoing diffusons and group the
remainder
of the diagrams into a skeleton $K$. Making use of Eq.\
(\ref{fluxladder2}) we obtain:
\begin{eqnarray} \label{inte} \langle T_a^j\rangle _{\rm
con}\!&\!= \!&\!\epsilon_a^j
(j\!-\!1)! \!
\int \! {\rm d}{\bf r}_1 {\rm d}{\bf r}_{1}'\cdots {\rm d}{\bf r}_j {\rm d}
{\bf r}_{j}'
{\cal L}^{\rm in}({\bf r}_1) {\cal L}^{\rm out} ({\bf r}_{1}')  \nonumber \\
&& \cdots {\cal L}^{\rm in}({\bf r}_j) {\cal L}^{\rm out}({\bf r}_{j}')
K({\bf r}_1, {\bf r}_{1}', \cdots ,
{\bf r}_j,{\bf r}_{j}').\end{eqnarray}
The integral just
describes \begin{equation} \langle{\rm Tr} (t t^\dagger)^j\rangle \equiv
\sum _{a_1,b_1,\cdots,a_j,b_j}\langle t_{a_1b_1}t^\dagger_{b_1a_2}
t_{a_2b_2}\cdots t_{a_jb_j}t^\dagger_{b_ja_1}\rangle
.\end{equation}
Indeed, for this
quantity
there is only one way to attach incoming and outgoing diffusons to $K$.  The
sums over the indices lead exactly to the total-flux diffusons
in Eq.\ (\ref{inte}). We thus find \begin{equation} \langle
T_a^j\rangle_{\rm
con}=(j-1)! \epsilon_a^j \langle {\rm Tr} (t t^\dagger)^j\rangle,
\end{equation} which is the
crucial step in the derivation. The eigenvalues $T_n$ of the
transmission matrix $t^\dagger t$ can be expressed as $T_n=1/\cosh ^2 L
\gamma_n$. Under
very general conditions \cite{nazarov} the distribution of the Lyapunov
coefficients $\gamma_n$ is
uniform \cite{dorokhov,pendry,stoneboek}. This implies that:
\begin{equation} \langle {\rm Tr}
(t t^\dagger)^j\rangle= \langle \sum_{n=1}^N T_n^j\rangle=g\int_0^1\frac{{\rm
d}T}{2T\sqrt{1-T}} T^j .\end{equation}
Normalizing with respect to the average,
we introduce
$ s_a=T_a / \langle T_a\rangle$.
The generating function of the connected diagrams
is easily calculated
\begin{eqnarray} \Phi_{\rm con}(x)&\equiv&\sum_{j=1}^\infty
\frac{(-1)^{j+1} x^j}{j!}\; \langle s_a^j\rangle_{\rm con} \nonumber \\
&=&g\log^2\left(\sqrt{1+x/g}+\sqrt{x/g}\right).
\label{phiconpw}\end{eqnarray}
Since the cumulants are solely given by connected diagrams, the
distribution of $s_a$ follows as
\begin{equation} p(s_a)
=\int_{-i\infty}^{i\infty}\frac{{\rm d}x}{2\pi i} \exp
 \left[xs_a- \Phi_{\rm con}(x) \right]
\label{PsaPW}.\end{equation}
For $s_a$ near unity and large $g$ we can expand $\Phi$ up to order $x^2$,
recovering the Gaussian behavior found by Kogan {\it et al.}\cite{kogan}:
\begin{equation}
p(s_a)\approx \sqrt{\frac{3g}{4\pi} } \exp[ -\frac{3g}{4} (s_a-1)^2]
\end{equation}
The integrand in Eq.\ (\ref{PsaPW}) has a branch cut from $x=-g$ to
$x=-\infty$. For $s_a \le 0$
the contour can be closed to the right and $p(s_a)$
vanishes.
The shape for small $s_a$ (and large $g$) is dominated by a saddle point.
One finds essentially a log-normal growth: \begin{equation}
p(s_a) \sim \exp\left[ \frac{g}{4}-\frac{g}{4} \left( \log \frac{2}{s_a}
+\log \log \frac{2}{s_a} -1 \right)^2 \right].  \end{equation}
Also for large $s_a$ we can apply steepest
descent. Here one finds a simple exponential decay \begin{equation}
p(s_a)\approx \exp(-gs_a+g\frac{\pi^2}{4}), \qquad s_a \gg 1.\end{equation}
In
Fig.\ \ref{figPtaPW} we present the distribution (\ref{PsaPW}) for
some values of $g$. At moderate $g$ the deviation from a Gaussian is
clearly seen.

So far we have considered the case of an incoming plane wave. In optical
systems a Gaussian intensity profile is more realistic. For
perpendicular incidence
the incoming amplitude is $ \psi^{\rm in}({\bf r})= W^{-1}
\sum_a \phi(q_a) \psi_a^{\rm in} ({\bf r})$,
where $\psi_a^{\rm in} ({\bf r})$ is the plane wave of Eq.\
(\ref{planepsi}), and where \begin{equation}
\phi(q_a)=\sqrt{2\pi}\rho_0\exp(-\frac{1}{4}\rho_0^2q_a^2).\end{equation}
We consider the limit where the beam is much broader than the
sample thickness ($\rho_0\gg L$) but still much smaller than the transversal
size of
the slab ($\rho_0\ll W$). (A smaller beam diameter complicates the problem;
when incoming transverse momenta, which
are
of order $1/\rho_0$, become of the order of $1/L$, the diffusons will take a
different form\cite{t3prl}). Here the momentum dependence of the diffusons
can be neglected.
Due to integration
over the center of gravity, each diagram involves a factor
$A\delta_{\Sigma q,\Sigma q'}$.
In the $j^{\rm th}$ order term there occurs a factor
\begin{eqnarray} F_j&=&\frac{A}{A^{2j}}\sum_{q_1 q_{1}'\cdots
q_{j}q_{j}'} \phi(q_1)\phi^{\ast}(q_{1}')\cdots\phi(q_{j})\phi^\ast(q_{j}')
\delta_{\Sigma q,\Sigma q'}\nonumber \\&=&\int {\rm d}^2\rho \;
|\phi(\rho)|^{2j}.\end{eqnarray} For a plane
wave we have $|\phi(\rho)|=\sqrt{A}$,
and $F_j=A^{1-j}$. For our Gaussian beam we
obtain \begin{equation}
F_j=\frac{1}{j}\left(\frac{\pi\rho_0^2}{2}\right)^{1-j}.\end{equation} It is
thus convenient to identify $A_G=\frac{1}{2}\pi\rho_0^2$ with the
effective area of a Gaussian beam. As compared to the plane
wave case, the $j^{\rm th}$ order term
is smaller by a factor $1/j$ for a Gaussian profile.
This implies for the generating function of the connected diagrams
\begin{equation} \Phi_{\rm con}(x) = g\int_0^1 \frac{{\rm
d}y}{y}\log^2\left(\sqrt{1+\frac{xy}{g}}+\sqrt{\frac{xy}{g}}\right)
\label{phicong}
.\end{equation} For small $s_a$ (and large $g$) there is again a log-normal
saddle point. For large $s_a$ the dominant shape of the decay is given by the
singularity at $x=-g$ and again yields $p(s_a)\sim\exp(-gs_a)$. In
Fig.\ \ref{figPtaG} we present the distribution function for different
values of $g$.

By expansion of the generating function we recover the results for the second
and third cumulant, obtained previously by de Boer {\it et al.} \cite{t3prl}
\begin{eqnarray} \langle s_a^2\rangle_{\rm cum}&=&\frac{2}{3g},\qquad \langle
s_a^3\rangle_{\rm cum}= \frac{12}{5}\langle s_a^2\rangle^2_{\rm cum}
\end{eqnarray} for a plane wave. For a Gaussian beam we recover
\begin{equation} \langle s_a^2\rangle_{\rm cum}=\frac{1}{3g},\qquad \langle
s_a^3\rangle_{\rm cum}= \frac{16}{5} \langle s_a^2\rangle^2_{\rm
cum}.\end{equation} These last relations were confirmed
experimentally\cite{t3prl}.

We apply the same method for the distribution of the angular
transmission coefficient. In the plane wave situation the average reads
$\langle T_{ab} \rangle = \epsilon_a \epsilon_b g$. Let us count the
number of connected loopless diagrams that contribute to
$T_{ab}^j=t_{ab}t^\dagger_{ba}t_{ab}\cdots t_{ab}t^\dagger_{ba}$.
In this case all pairings into outgoing diffusons contribute.
This yields an extra combinatorical factor $j!$ in the
$j^{\rm th}$ moment: \begin{equation} \langle T_{ab}^j \rangle_{\rm con}=
j! (j-1)! \epsilon_a^j \epsilon_b^j \langle {\rm Tr} (t^\dagger t)^j\rangle.
\end{equation}
For the normalized angular transmission coefficient
$s_{ab}=T_{ab}/\langle T_{ab}\rangle$, we introduce the following
generating function of the connected diagrams: \begin{equation} \Psi_{\rm
con}(x)\equiv \sum_{j=1}^\infty \frac{(-1)^{j-1}x^j}{j! j!} \langle
s_{ab}^j\rangle_{\rm con}
\end{equation}
It is easy to see that
\begin{equation}
\Psi_{\rm con} (x)=\Phi_{\rm con} (x)
,\end{equation} with $\Phi_{\rm con}$ given by Eq.\ (\ref{phiconpw}) for
plane wave incidence and by Eq.\ (\ref{phicong}) for a broad
Gaussian beam, respectively.
In contrast to the total-transmission distribution, the cumulants are not
only given by the connected diagrams. Kogan {\it et al.} showed that the
summation of the disconnected diagrams can be done elegantly
by performing
an additional integral. Using this result one gets \begin{equation}
p(s_{ab})=\int_0^\infty \frac{{\rm d}v}{v}\int_{-i\infty}^{i\infty}
\frac{{\rm d}x}{2\pi i} \exp\left(-\frac{s_{ab}}{v}+ xv-\Psi_{\rm con}(x)
\right) \label{PtabPW}.\end{equation} The speckle intensity distribution is
plotted in Fig.\ \ref{figPtabP} for an incoming plane wave.

For large $g$ and moderate
$s_{ab}$ we have $\Phi_{\rm con}(x) \approx x $ and we recover the Rayleigh
law: $p(s_{ab})=\exp(-s_{ab})$.
The leading correction is found by expanding in $1/g$
\begin{equation}
p(s_{ab})= {\rm e}^{-s_{ab}} \left[ 1+\frac{1}{3g} (s_{ab}^2 -4 s_{ab}+
2)\right] \end{equation}
This was derived previously by Shnerb and Kaveh\cite{shnerb}; here we have
related
the prefactor of the correction term to the conductivity. Genack and Garcia
fitted their data to this relation and found $g=14.6$. Our Eq.\
(\ref{PtabPW}) describes these data very well for $g=14.4$.

For large $s_{ab}$ one can again apply steepest descent, which yields
\begin{equation} p(s_{ab})\sim \exp(-2\sqrt{g s_{ab} }\;). \end{equation}
This stretched
exponential tail differs from the form
$p(s_{ab})\sim\exp[-(81 g s_{ab}^2 /16)^{1/3}]$ asserted by Kogan {\it et
al.} Their findings are based on truncating $\Phi_{\rm con}(x)$ after order
$x^2$, which corresponds to including only the simplest connected diagram.
Taking the full generating function into account, we find a qualitatively
different saddle point.

A Gaussian profile of the incoming beam leads to a different distribution
with the same asymptotic behavior.

Our results remain valid if there is a mismatch in the indices of
refraction\cite{thmn5}, but may change if the system is optically not very
thick.

In conclusion, we have calculated the distribution function of the total
amount of light transmitted through a disordered slab. It was related to the
distribution function of the eigenvalues of the transmission matrix, which is
known since the Lyapunov coefficients are uniformly distributed. Comparing to
previous diagrammatic calculations \cite{t3prl} it became clear that this
uniform distribution comes from summing connected, loopless diagrams.

Next we have calculated the distribution of the angular transmission
intensity. Here a deviation from Rayleigh statistics is found, of new
stretched exponential form. Our calculations are in agreement with
measurements of both total and angular resolved transmission. As our results
are based on loopless diagrams, they constitute the mean field expressions.
Near the Anderson transition loop diagrams will
change the distribution functions.

\acknowledgments Discussions with I.~V. Lerner, B.~L. Altshuler, and Yu. V.
Nazarov are gratefully acknowledged. The research of Th.~M.~N. was
supported by the Royal Netherlands Academy of Arts and Sciences (KNAW). This
work was also sponsored by NATO (grant nr. CGR 921399).

\newpage
\begin{figure}
\caption{Intensity distribution of the total transmission, in units of
$\protect\sqrt{3g/4\pi}$,
versus the normalized intensity $s_a$ for an incoming
plane wave. With $ g=2,4,16$, and 64 (upper to lower curve). }
\label{figPtaPW}
\end{figure}

\begin{figure}
\caption{Intensity distribution of the total transmission, in units of
$\protect\sqrt{3g/2\pi}$, versus
the normalized intensity $s_a$ for an incoming wave with Gaussian profile.
With $g=2,4,16$, and 64 (upper to lower curve).}
\label{figPtaG}
\end{figure}

\begin{figure}
\caption{Intensity
distribution of speckles versus the normalized intensity $s_{ab}$ for an
incoming plane wave. With $g=2,4$, and 8 (upper to lower curve at
$s_{ab}=5$). The dashed line corresponds to
the Rayleigh law ($g= \infty$).} \label{figPtabP}
\end{figure}

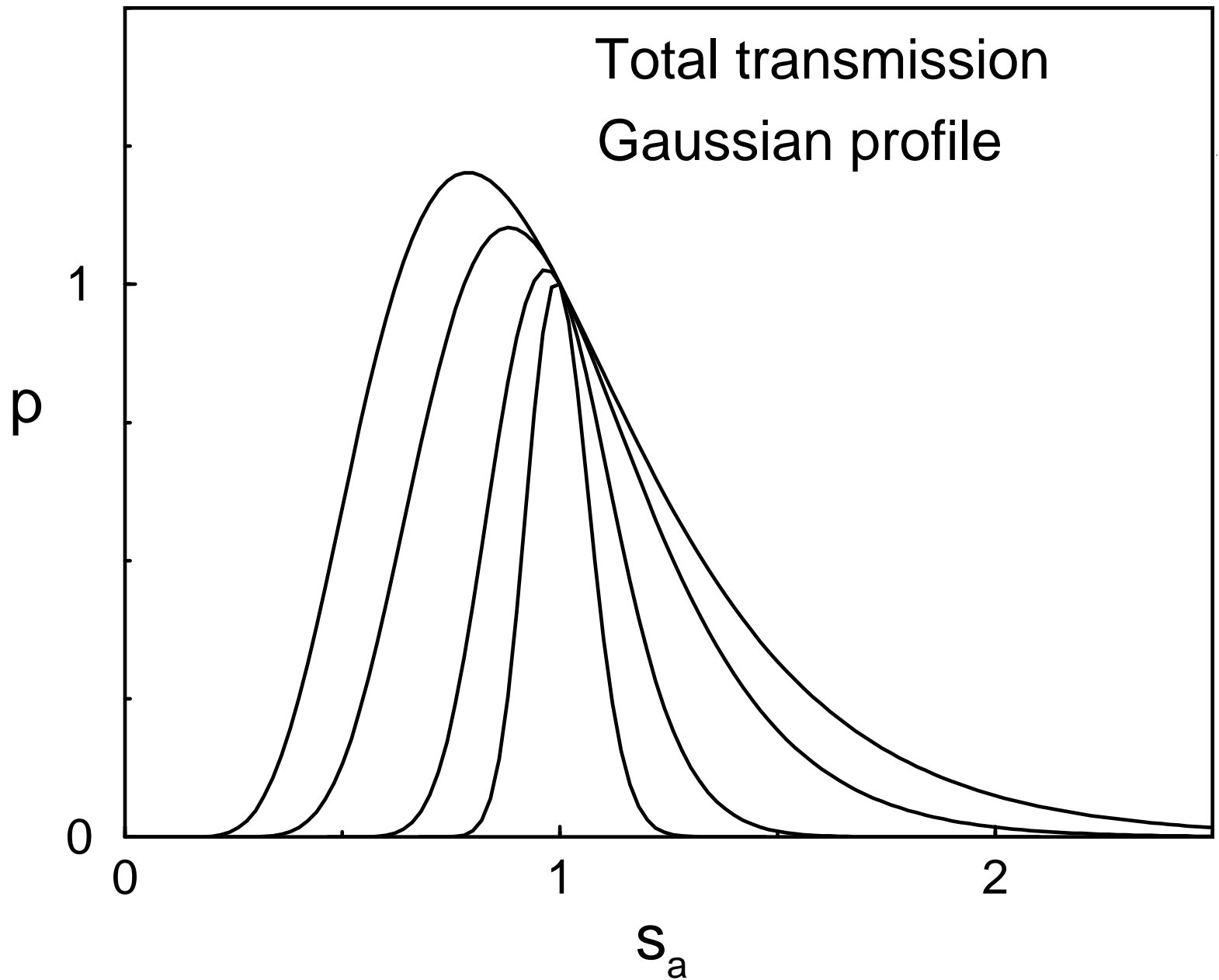
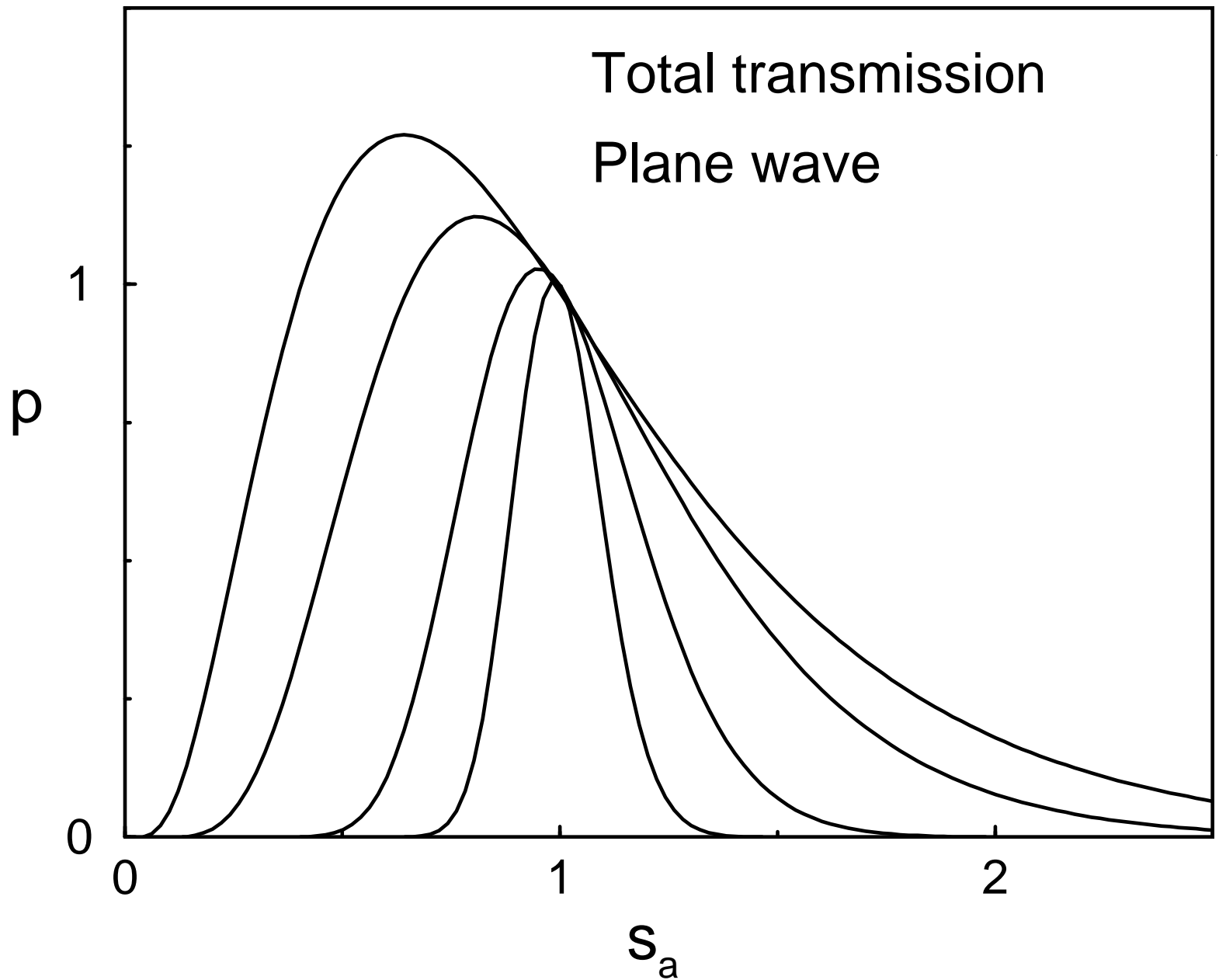
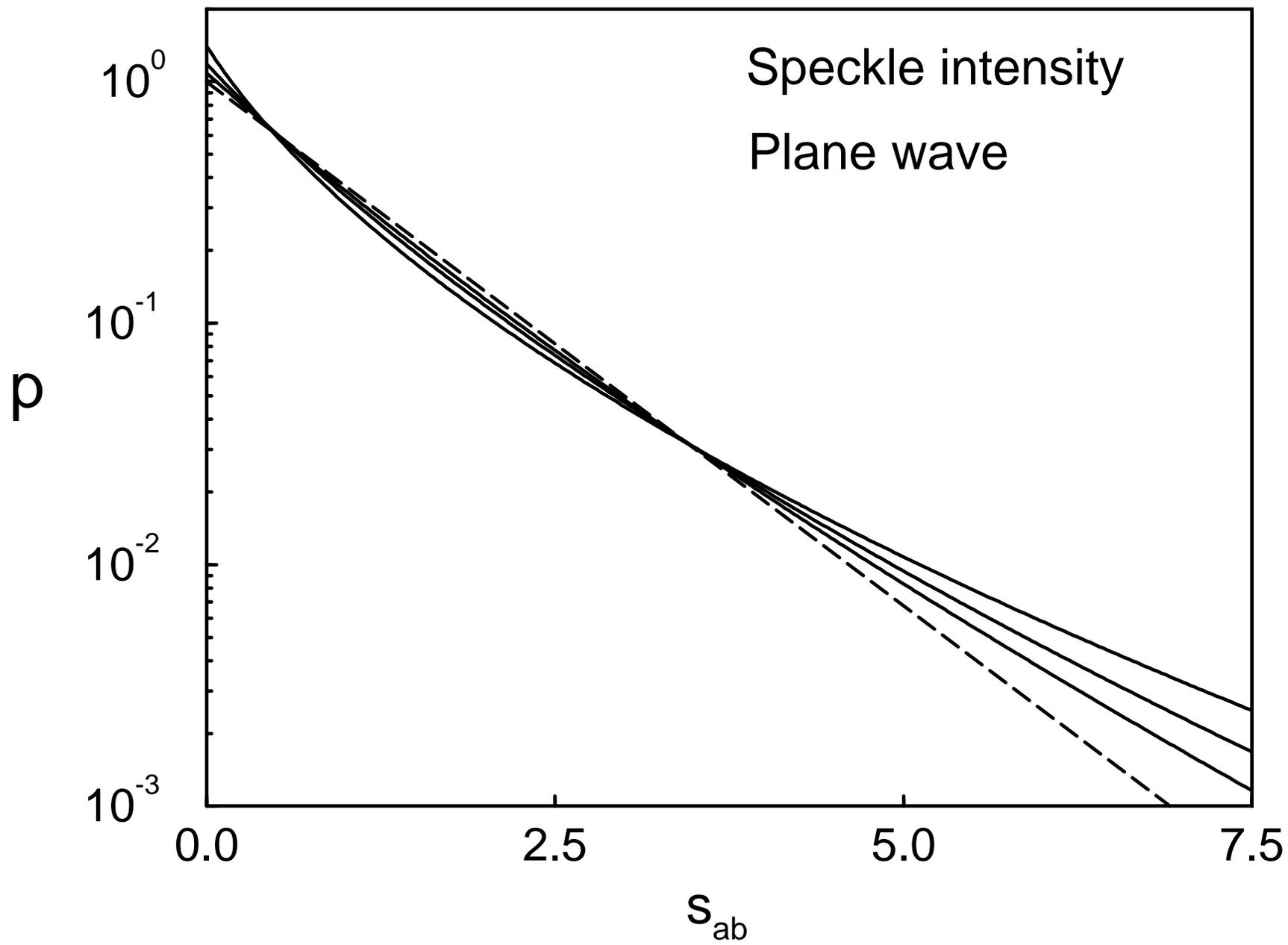
\end{document}